\documentclass[aps,prm,showpacs,reprint]{revtex4-1}
\usepackage{epsfig}
\usepackage{dcolumn}
\usepackage{epstopdf}
\usepackage{graphicx}
\usepackage{ulem}
\usepackage{color}
\usepackage{bm}
\begin{document}
\title{Efficient thermoelectricity in Sr$_2$Nb$_2$O$_7$ with  energy-dependent relaxation times}
\author{Giulio Casu}\author{Andrea Bosin}\author{Vincenzo Fiorentini}
\affiliation{Department of  Physics, University of Cagliari, Cittadella Universitaria, I-09042 Monserrato (CA), Italy}
\date{\today}

\begin{abstract}
We evaluate theoretically the thermoelectric efficiency of the layered perovskite Sr$_2$Nb$_2$O$_7$ via calculations  of the electronic structure and transport coefficients within  density-functional theory and Bloch-Boltzmann relaxation-time transport theory.  The predicted figure-of-merit tensor $ZT$, computed with energy-, chemical potential- and temperature-dependent relaxation times, has one component increasing monotonically from around 0.4 at room temperature to 2.4 at 1250 K at an optimal carrier density around 2$\times$10$^{20}$ cm$^{-3}$, while the other components are small. The Seebeck coefficient is about 250 to 300 $\mu$V/K at optimal doping, and reaches 800 $\mu$V/K at lower doping. We provide a  {\tt python} code  implementing various approximations to the energy-dependent relaxation time transport, which can  be used to address different systems with an appropriate choice of material parameters. 
\end{abstract}
\maketitle

\section{Introduction}
Thermoelectricity as  an energy source has been the focus of much research effort recently.  Materials are assessed as candidate thermoelectrics  based on their   figure of merit 
$$
{ZT}=\frac{\ \sigma S^2}{\kappa_e +\kappa_L} T,\nonumber
$$
with $\sigma$ the electrical conductivity, $S$ the Seebeck coefficient, $T$ the temperature, and $\kappa_e$, $\kappa_L$ the electronic and lattice   thermal conductivities.  Recently, we  analyzed theoretically a few  materials \cite{noi2,LTO}, and in particular \cite{LTO} the layered-perovskite La$_2$Ti$_2$O$_7$ (LTO), finding interesting figure-of-merit results. In this paper, we study  the layered perovskite Sr$_2$Nb$_2$O$_7$ (henceforth SNO), which is  fairly close  to LTO in terms of  low thermal conductivity, and has a well established growth procedure. The  predicted figure of merit of SNO is very good,  monotonically increasing from 0.3 at room temperature to  2.4 at 1250 K. 

We also  examine  several approximations to the definition and use of relaxation times in 
the  transport coefficients calculations in the Bloch-Boltzmann approximation \cite{allen}, extending  our 
approach in Ref.\cite{LTO}. We provide and discuss a {\tt python} code \cite{thermocasu} that 
performs the calculations using data generated by the ab initio {\tt VASP} \cite{vasp} code, and using the 
{\tt BoltzTrap2} \cite{bt2} code as a library. 

\begin{figure}[ht]
\centerline{\includegraphics[width=0.5\linewidth]{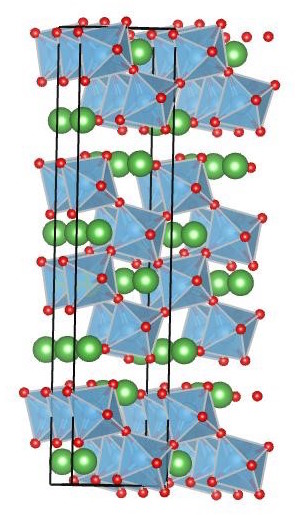}}
\caption{\label{struc} Sketch of the structure of SNO. The $a$ axis points into the page, the $b$ axis vertically, and the $c$ axis from left to right. The primitive cell is outlined.}
\end{figure}

\section{Method and auxiliary results}

\subsection{General}

The ingredients of $ZT$ are the electronic transport coefficients  that  can be  obtained from the electronic structure (electrical and electronic-thermal conductivity, Seebeck thermopower),  and  the lattice thermal conductivity. For the latter, we adopt the experimental data of Ref.\cite{SNO-therm}: about 1 W/K/m, temperature independent, along the $b$ axis; 1.8 W/K/m, roughly isotropic and temperature independent, in the $a$,$c$ plane. This assumption is justified by the fact that, based on our previous analysis of the thermal conductivity (Ref.\cite{LTO}, Sec.IIC) in the analogous material La$_2$Ti$_2$O$_7$, the low thermal conductivity of SNO is very probably  intrinsic to the material, and not defect-originated.

For the electronic transport coefficients  we  use the ab initio density-functional band structure to calculate the coefficients as function of temperature and doping in  the relaxation-time approximation to the linearized  Boltzmann transport equation, known as Bloch-Boltzmann theory \cite{allen,bt2}. We explore several approximations to  the energy- and temperature-dependent relaxation time (Secs. \ref{trcoe} and \ref{tauappr}).
Based on preliminary calculations and the behavior of LTO, we choose to concentrate on $n$-type doping. 

\subsection{Ab initio calculations}\label{subsec:abInitioCalculations}

Ab initio density-functional structure optimization and band-structure calculations are performed with the generalized-gradient approximation \cite{pbe}  and the projector augmented wave method \cite{paw} using the {\tt VASP} code \cite{vasp} with  {\tt \verb+Sr_sv+}, {\tt \verb+Nb_pv+}, and {\tt \verb+O_s+} PAW datasets, using the  maximum of all  suggested maximum cutoffs, 283 eV (increased as usual by 30\% for calculations involving stress). The  structure of the orthorhombic phase of SNO (Fig.\ref{struc}) has space group {$Cmc2_1$} up to 1615 K, above which it becomes $Cmcm$ (we  neglect the minor incommensurate  modulation \cite{ohi} below 490 K as it preserves closely the layered structure and the space group). We optimize the structure  following quantum forces (threshold 0.01 eV/\AA) and stress (threshold 0.5 kBar). The  computed lattice constants are $a$=4.00 \AA, $b$=27.30 \AA, $c$=5.82 \AA, which are 1.5\%, 2.1\%, and 2.4\% from the experimental values \cite{SNO-therm}  3.933, 26.726, and 5.683 \AA, as expected due to the density functional we use. Electronic states are calculated  on a (24$\times$8$\times$16) k-points grid. 

\begin{figure}[ht]
\centerline{\includegraphics[width=1\linewidth]{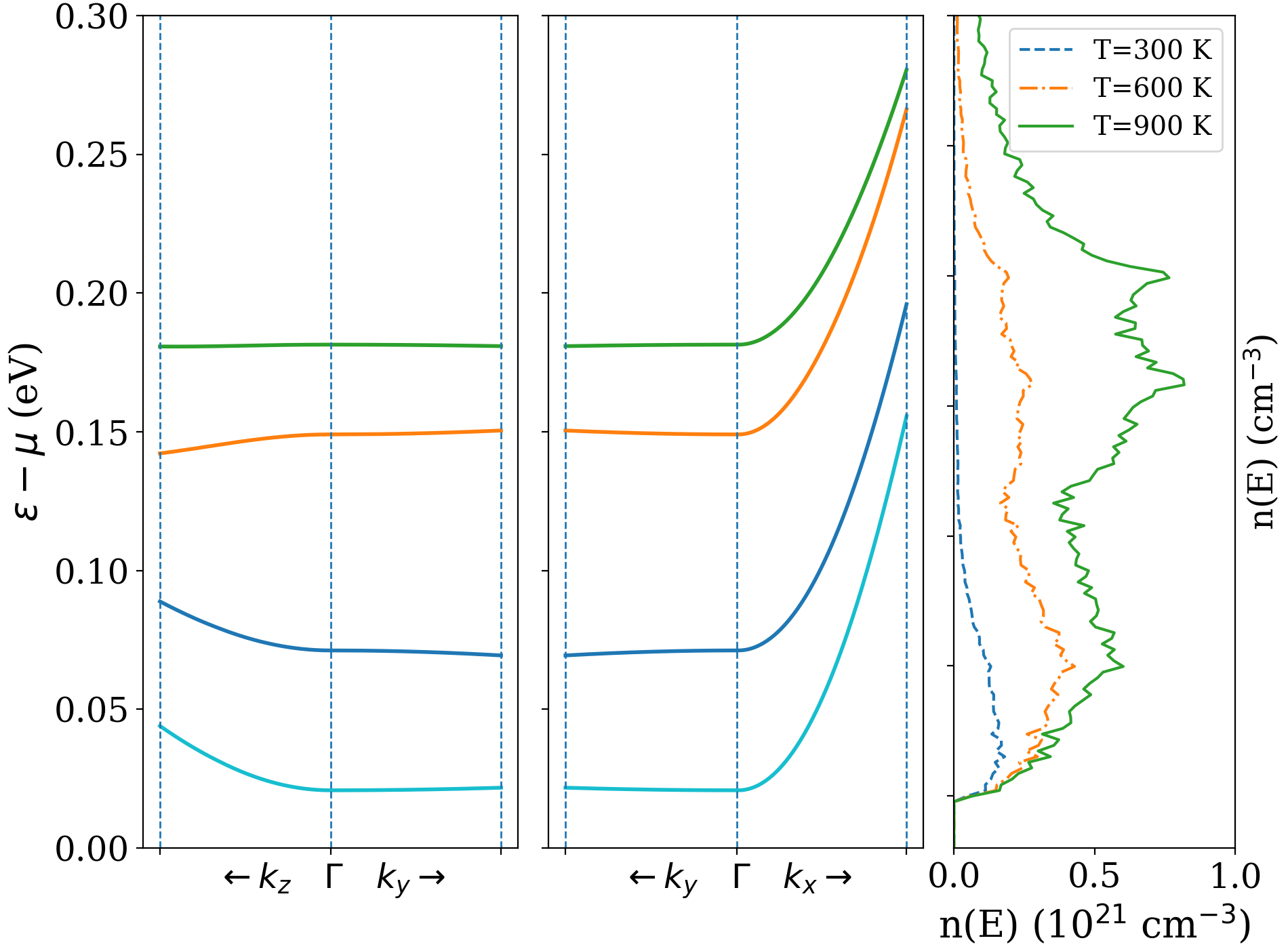}}
\caption{\label{bandfermidos} Left, center: conduction bands of SNO along the three reciprocal lattice directions, starting at their minimum at $\Gamma$. Right: energy-dependent carrier density, i.e.  Fermi dis\-tri\-bution times the density of states, for $T$=300, 600, 900 K. The chemical potential is 15 meV below the conduction edge.}
\end{figure}

The conduction band minimum is at $\Gamma$, while the valence band maximum is at $X$=($\pi$/$a$,0,0); the minimum gap is 2.5 eV (underestimated, as expected, in comparison to roughly 4 eV experimentally \cite{ohi2,elda}), so the $n$-type coefficients are essentially   unaffected by valence states at any temperature. A close-up view of the conduction bands in the vicinity of the $\Gamma$ point is in Fig.\ref{bandfermidos}, left and center panel; the rightmost panel shows the density of states weighed by the Fermi-Dirac function  (i.e. the energy-dependent carrier density), which clearly comes from the first four conduction bands at typical temperatures. (The density of states is recalculated by our code,   Sec.\ref{thecode}, from the interpolated band structure on a 120$\times$40$\times$80 k-grid.)

\subsection{Transport coefficients and  relaxation time}
\label{trcoe}
   To compute the transport coefficients accounting for energy-dependent relaxation times we built a new  code \cite{thermocasu} (see Sec.\ref{thecode})
using parts of the  {\tt BoltzTrap2} (BT2 henceforth) \cite{bt2} transport code as libraries. Via BT2, the ab initio bands (assumed rigid, i.e. not changing with doping or temperature) are interpolated as explained in Ref.\cite{bt2} over  a k-grid much finer than the ab initio one. For the calculations reported below, the grid contains 60 times more points than the ab initio grid, so it is approximately equivalent to a (94$\times$32$\times$62) grid.

As discussed in Refs.\cite{noi2} and \cite{LTO}, in the constant-relaxation time approach the constant $\tau$=$\tau_0$ will factor out of the integrals determining the Onsager coefficients. In such case, the BT2 code returns the reduced coefficients $\overline{\sigma}_0$=$\sigma$/$\tau_0$ and $
\overline{\kappa}_{e,0}$=$\kappa_e$/$\tau_0$,  determined  by the band 
structure, temperature and doping, but independent of $\tau_0$.  $ZT$ is then calculated from the reduced transport coefficients under some hypothesis for the relaxation time $\tau_0$. This is   generally a poor approximation because a constant $\tau_0$ is  uncertain and largely arbitrary, and it cannot account for various relevant physical effects, such as  Fermi distribution tails, phonon occupation changes, and more. 
 
 To improve over the constant-time approximation, we 
 adopt a model     relaxation time which depends on temperature $T$, chemical potential $\mu$, and carrier energy $E$. $\tau$ enters the kernel  of the Fermi integrals (see Eq.9 of Ref.[3]) which provide the building blocks of the Onsager transport coefficients. The relaxation time is
$$\tau(T,E,\mu)=1/P_{\rm imp}+1/P_{\rm ac}+1/P_{\rm polar}$$
where  the rates $P$ of impurity, acoustic-phonon and polar-phonon scattering are given in Refs.\cite{noi2,ridley1,ridley2} and, for convenience, in the Supplementary Material \cite{smat}. Piezoelectric  scattering is neglected because the relevant  matrix element \cite{ridley1} is zero by symmetry \cite{nye,electrom}. Also, as in LTO, spontaneous polarization points along the $c$ axis, so  by symmetry it will not affect transport along $a$; as it turns out,  thermoelectric transport in SNO is sufficiently interesting only along the $a$ axis.

  \begin{figure}[ht]
\centerline{\includegraphics[width=1\linewidth]{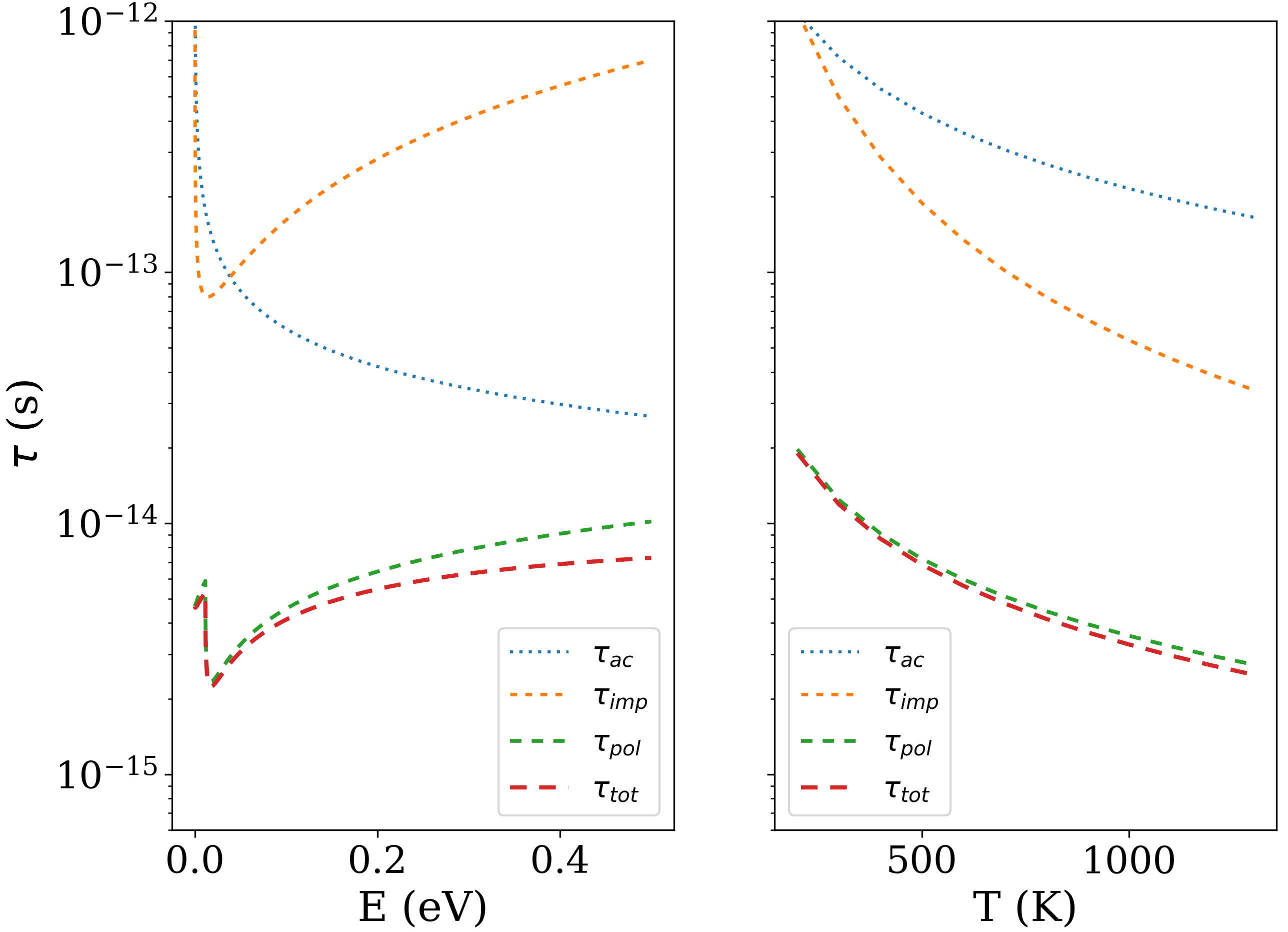}}
\caption{\label{taufig1} Relaxation time $\tau$($T$,$E$) vs $E$ ($T$=700 K) 
and $T$ ($E$=54 meV above the conduction edge). $\mu$ is fixed at zero.}
\end{figure}

The behavior of $\tau$($E$,$T$) in SNO  is sketched vs $E$ and $T$ in Fig.\ref{taufig1}. Clearly the LO polar phonon scattering dominates, and its downward jump across the LO phonon energy is at very low energy,  the most relevant region for transport (the detailed parameter values involved are discussed below). 
We mention in passing that it is now becoming possible \cite{giustino} to compute  ab initio the full k and energy  dependence of the $\tau$ components related to electron-phonon scattering. This approach is, however, hugely  more complex than ours and is still in its infancy in the field of thermoelectricity, and is therefore beyond our present scope.

\subsection{Approximations and parameters for $\tau$}
\label{tauappr}
With the $\tau$ model just discussed,  here  we calculate the  transport coefficients $ZT$ using two  approximations. The best one is the  NCRT (non-constant relaxation time), which uses the full energy-dependent time in the calculation of the Fermi integrals, including its chemical-potential dependence (specifically $\tau$ multiplies the kernel $\sigma$ in the integral of Eq.9, Ref.\cite{bt2}. 
The  ART (average-relaxation-time) approximation uses instead  the $T$- and $\mu$-dependent, energy-averaged \cite{cardona} 
 $$
\tau_{\rm ART}(T, \mu)=
\frac{2}{3}\frac{\int_0^{\infty} \, E\ \, \tau(T,E,\mu)  D(E)(-\frac{\partial f(T, E,\mu)}{\partial E})\, dE}{\int_0^{\infty}
 D(E) f(T, E,\mu)\, dE}
$$
($D$($E$) is  the density of states, and  $f$  the Fermi-Dirac distribution) to compute $ZT$ vs $T$  from the reduced  transport coefficients,  as in Eq. Ref.\cite{LTO}, Eq.3. The relaxation time, computed off-line and only once for each $T$ and $\mu$, still accounts for the $T$ dependence of $\mu$, Fermi function, etc., also improving slightly over Ref.\cite{LTO}, which used a parabolic band density-of-states.  

 NCRT is typically relatively inexpensive (depending on the system and control parameters, up to at most a few hours on a laptop with a few Gbytes memory usage, but typically much less than that). The ART is even  lighter, and is useful for exploratory work, providing in our experience a good guide to the  full  results (see the discussion below).  Of course, constant relaxation-time results (labeled CRT below) can be compared recalculating the reduced coefficients with BT2 and using  a constant time (which we  choose to be $\tau_0$=5 fs) to obtain $ZT$. All three codes are provided in the Supplementary Material \cite{smat} as well as on line,  and discussed in Sec.\ref{thecode}. 

%
The model $\tau$ requires several parameters. Some  are imported from experiment or previous calculations:  dielectric constants $\varepsilon_{\infty}$=4.7  
and  $\varepsilon_{\rm lattice}$=33.5 \cite{epsinf,epsil}; average sound velocity  $v$=4438 m/s \cite{SNO-therm};  dominant LO-phonon  frequency $\hbar$$\omega_{\rm LO}$=11.4 meV \cite{discLO}. Some others are computed directly:  effective conduction mass $m_c^*$=0.3 $m_e$, deformation potential $D$=10 eV, density 4930 kg/m$^3$. The mass is obtained from a polynomial fit to the conduction band vs k (we use the $a$ component of the mass tensor, since the transport along the other directions is negligible anyway); the deformation potential is a numerical derivative of the conduction edge energy vs strain. We note that the lowest LO phonon with $a$-polarization in STO is much lower than in LTO, which enhances the polar scattering at  low energy. We  use only one phonon replica in the polar scattering rate. 
  
\section{Results}

\begin{figure}[ht]
\includegraphics[width=1\linewidth]{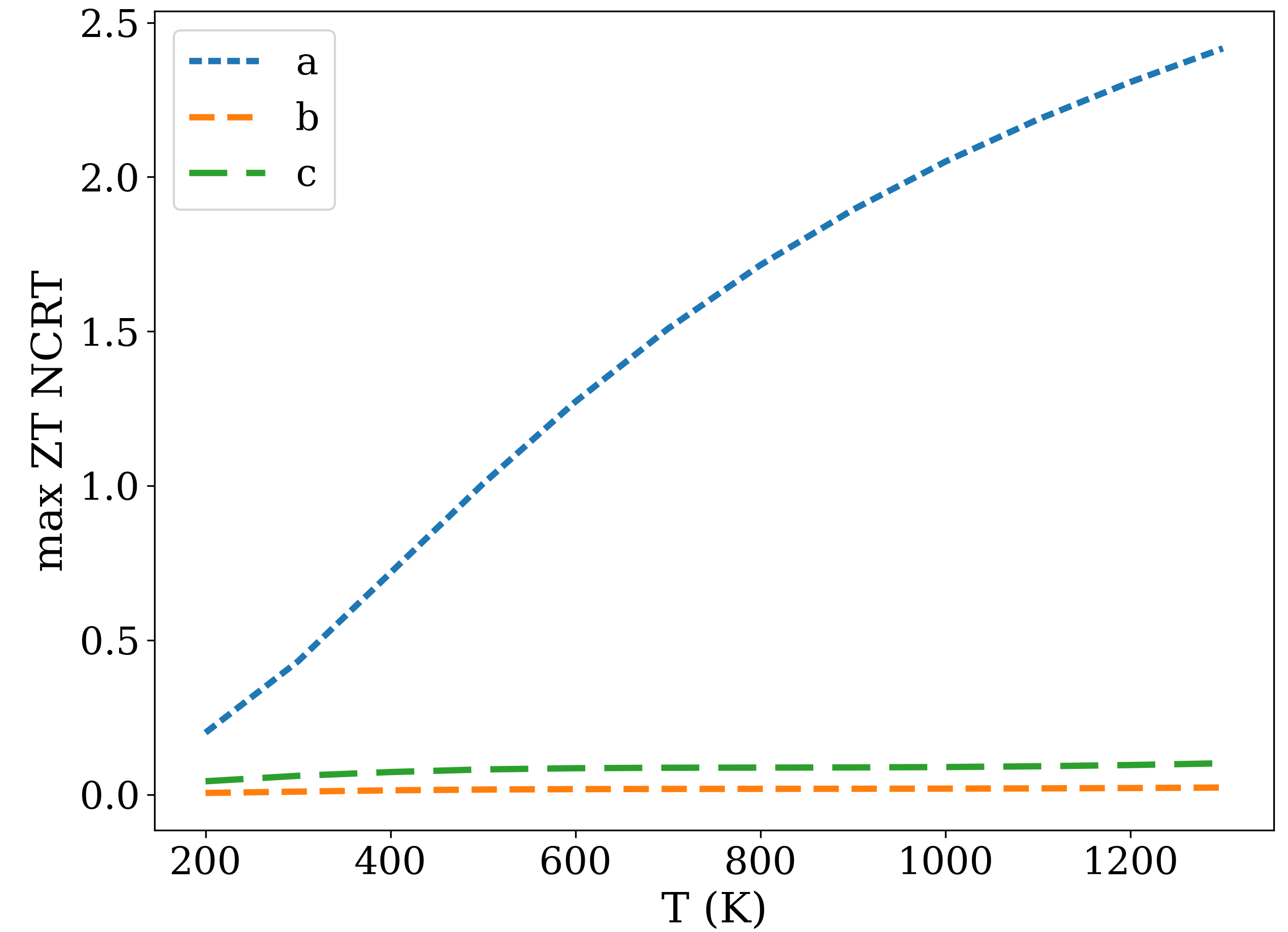}
\caption{\label{fig5} Components of the $ZT$ tensor at optimal doping (off-diagonal components are zero).}
\end{figure}

In Fig.\ref{fig5} we show the diagonal components of the $ZT$ tensor at optimal doping (i.e. the maximum value as function of doping; as shown below, in this case the optimal doping changes little with temperature) obtained with our best approximation NCRT. The $b$ and $c$  components  are small and of no interest for thermoelectricity, so we  disregard them henceforth.   The $a$ component of $ZT$ is instead large and therefore of practical interest, especially in the 500-1000 K range, where it is between about 1 and 2.  

\begin{figure}[ht]
\includegraphics[width=1\linewidth]{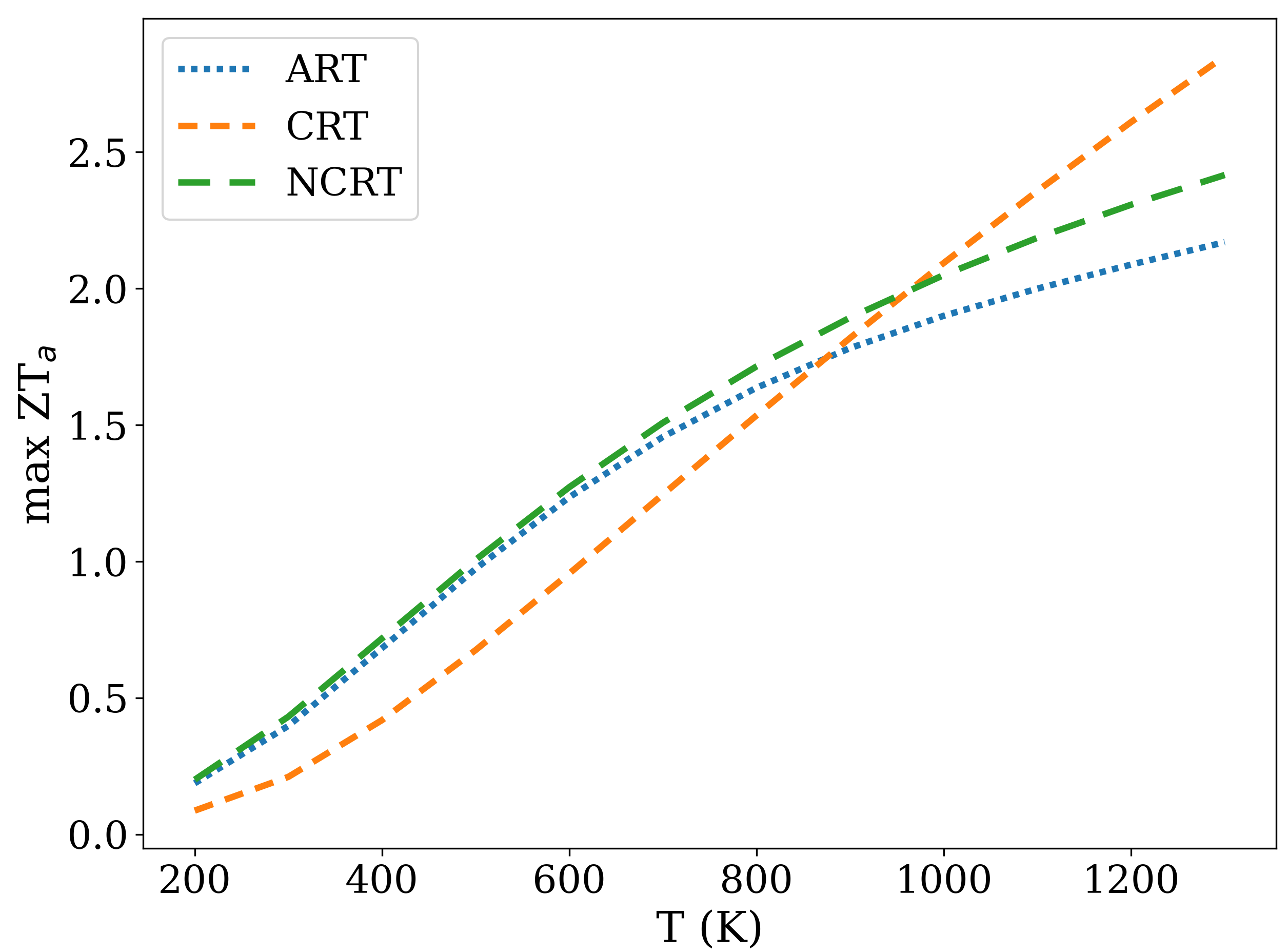}
\caption{$ZT$ vs $T$ ($a$ component) for the different  relaxation time treatments. In CRT, $\tau_0$=10 fs.}
\label{fig6}
\end{figure}

In Fig.\ref{fig6} we show $ZT$ ($a$ component) vs  $T$ at optimal doping  for the three methods discussed above. Clearly, there are no dramatic deviations between ART and NCRT. The CRT result, aside from its direct dependence on $\tau_0$ (which in this case was chosen sensibly after the fact), deviates most from the others due to its lack of temperature dependence.

\begin{figure}[ht]
\includegraphics[width=1\linewidth]{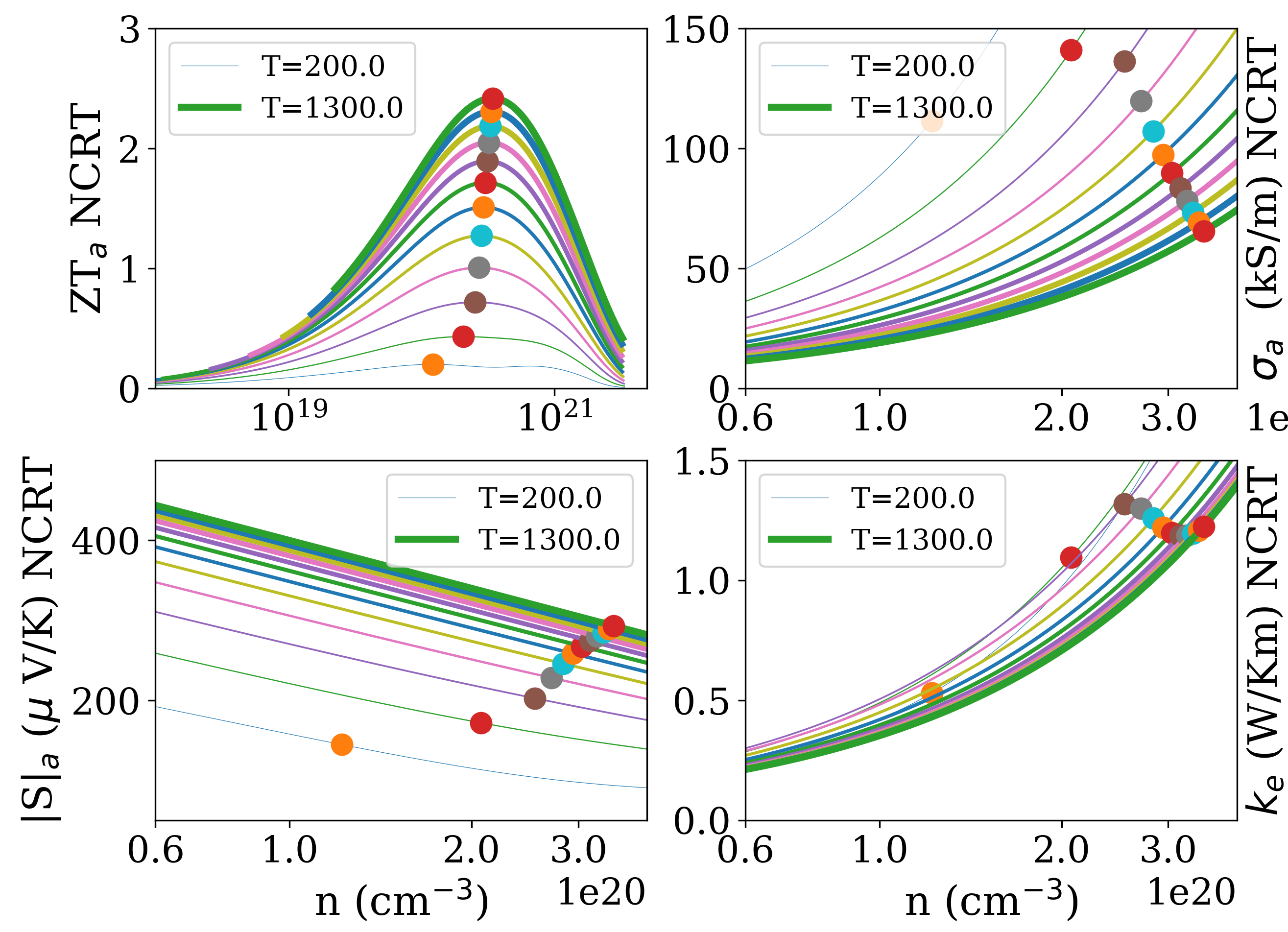}
\caption{$ZT$, Seebeck, conductivity, and electronic thermal conductivity  vs doping  at different $T$ in the NCRT approach. $T$ increases from 200 to 1300 K in steps of 100 K, indicated by increasing thickness. Dots indicate the values at optimal doping, i.e. where $ZT$ is maximal at that $T$.}
\label{fig7}
\end{figure}

Next,  in Fig.\ref{fig7} we report $ZT$, Seebeck, electrical conductivity, and electronic thermal conductivity vs doping as obtained by NCRT, for $T$ between 200 and 1300 K. The different curves  for increasing $T$ have increasing line thickness.  The dots indicate the values at optimal doping, i.e. the doping where ZT is maximal, which is  between 1 and
3$\times$10$^{20}$ cm$^{-3}$.
 Of course, it is easy to read off the graph any of the quantities at a fixed doping;  for example $ZT$$\simeq$1 
at 5$\times$10$^{19}$ cm$^{-3}$ and 900 K. Similarly the $T$ behavior can be inferred following the position of the dots, and it is similar to that of LTO \cite{LTO}.

This prediction of $ZT$ vs doping is  also, implicitly, a prediction of the optimal doping needed to achieve maximum $ZT$.  We are in  no position to evaluate the dopability of SNO, so it  remains to be seen if optimal doping, or anything  close to it, can be achieved experimentally. 
Also, we recall in passing that the conductivities are determined solely by electrons living in the  perfect-crystal bands and subject to  scattering from (mostly polar) phonons and charged impurities as embodied in $\tau$($T$,$E$,$\mu$), but no  scattering is accounted for from  disorder, dislocations, neutral impurities and traps, etc; these  could affect transport in ways  we cannot quantify.

In this context, we note that the material may be useful even if not highly dopable, for applications requiring just large thermoelectric power: $|S|$  is indeed larger at low doping, about over 750-800 $\mu$V/K at  10$^{18}$ cm$^{-3}$.

\begin{figure}[ht]
\includegraphics[width=1\linewidth]{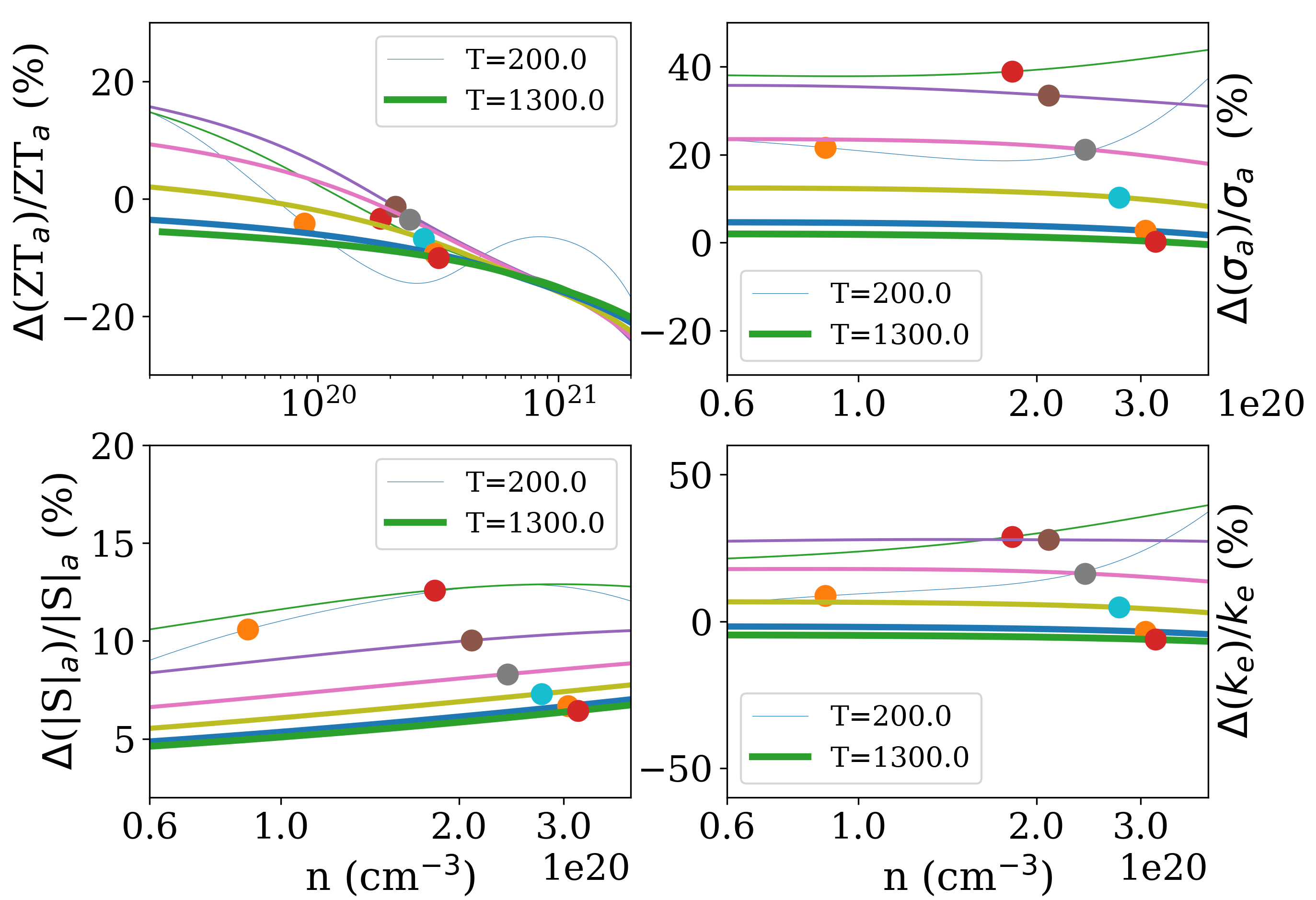}
\caption{Percentage deviations of  the various quantities calculated with ART from  those  in NCRT in Fig.\protect\ref{fig7}. Dots as in Fig.\ref{fig7}}
\label{fig8}
\end{figure}

In Fig.\ref{fig8} we report ($X_{\rm ART}-X_{\rm NCRT}$)/$X_{\rm NCRT}$, the relative deviation  between the ART and NCRT methods for quantity $X$. It appears that ART is within about $\pm$5 to 10\% of the full result for $ZT$ in the optimal doping region. The conductivities are  overestimated (by 0 to roughly 40\%) in ART, although the  difference largely  compensates in the $ZT$ ratio.  Interestingly, this Figure also reveals the extent to which the treatment of $\tau$ affects the Seebeck coefficient.  $S$  does depend on $\tau$ in NCRT, whereas by construction it does not in ART. The  NCRT description appears to reduce $S$ (in absolute value) by about 5 to 10\% in this material. Note that, of course, none of the above findings  need be generally  applicable to other systems.

  \begin{figure}[ht]
\centerline{\includegraphics[width=1\linewidth]{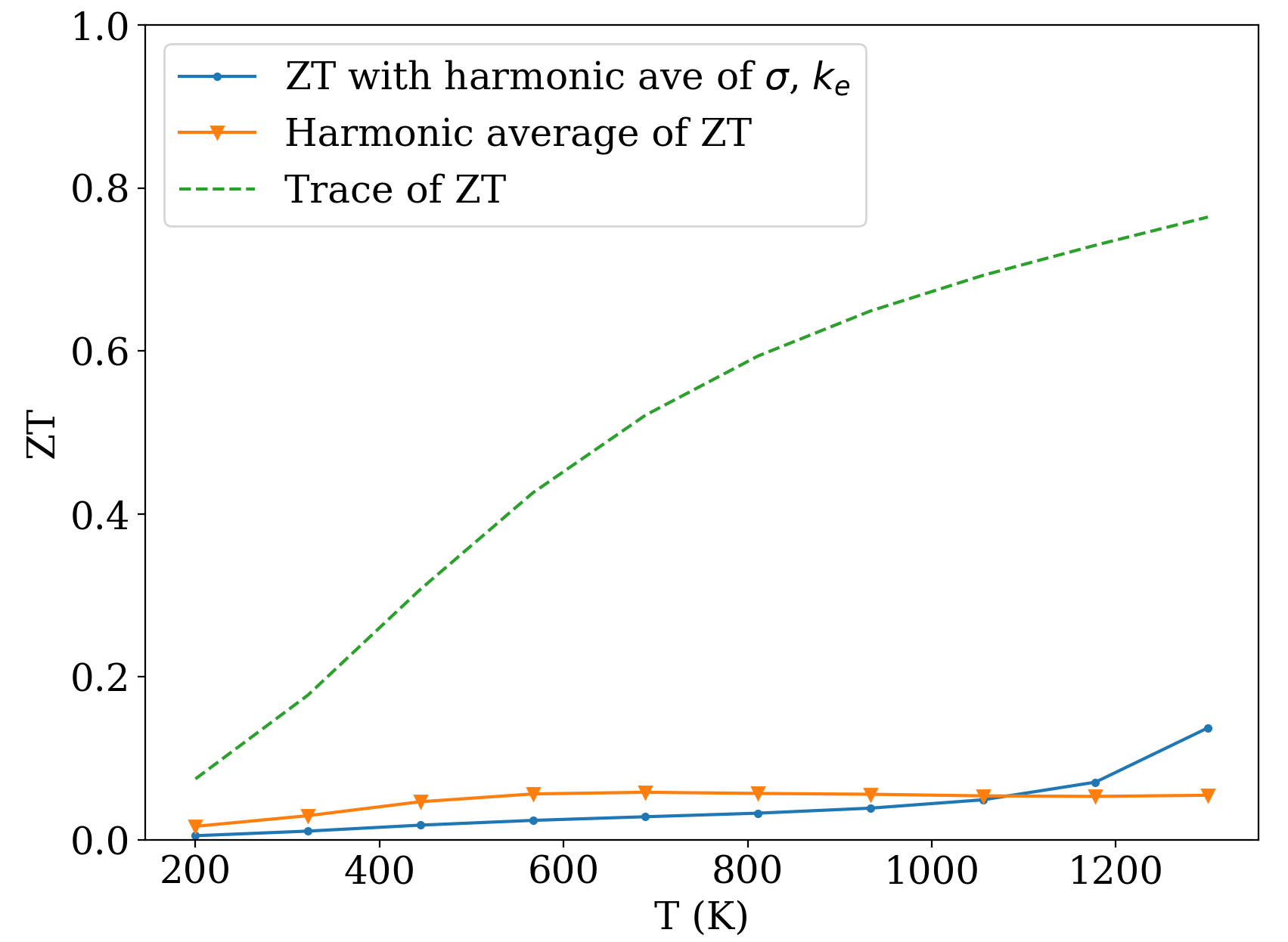}}
\caption{\label{trace} Three types of directional averages of $ZT$.}
\end{figure}

We close with an estimate of a directional average of $ZT$, potentially useful for polycrystalline material. 
In Fig.\ref{trace} we report three averages: the trace of $ZT$, the harmonic average \cite{harm} of the components of $ZT$, and $ZT$ calculated from the harmonic averages of the conductivities and the trace of the Seebeck. The latter  is probably the appropriate choice, but in any event the results are, as expected, much less exciting due to  the small $c$ and $b$ components and the dominance of small conductivities in the averages. This suggests that the scope for SNO  in thermoelectricity is  as a monocrystal, not quite as a polycrystal. This may not be entirely disheartening, because  the structural and growth properties \cite{locq} of SNO render it unlikely to be produced in the form of 
nanometer-sized polycrystals, multi-micron-sized crystallites as in sintered powder ceramics being
likely \cite{ceram}). Decently sized monocrystals are indeed fairly common and should be the ultimate goal of growth experiments.

\section{The code, and HOW TO USE IT}
\label{thecode}

Our code \cite{thermocasu}  is written in {\tt python} (tested up to version 3.7) and is free and open source. 
 We  provide  its current snapshot in the Supplementary Material \cite{smat}, with the {\tt vasprun.xml} and {\tt input} files used for the present material. The code comes in three stand-alone  versions   implementing the  approximations CRT, ART, NCRT discussed above. 

In  CRT and ART,  the reduced
coefficients  
calculated by  BT2 \cite{bt2} are used to compute $ZT$ with either the constant $\tau_0$ or, in the case of 
ART, with an energy-averaged  $\tau$, computed separately.
If NCRT is used, the code follows BT2 \cite{bt2} up to the calculation of
 energies $\varepsilon_b$({\bf k}) on the fine grid, at which point it computes the  relaxation time $\tau$($\varepsilon$) on the same energy grid, and  multiplies the integrand of the Fermi 
integrals by $\tau$. After that, it goes back to the BT2 sequence and eventually outputs the full coefficients. The current version works for $n$-type doping;  extension to $p$-type may be implemented, in which case the code will be updated on \cite{thermocasu}. 

\begin{figure}[ht]
\includegraphics[width=0.9\linewidth]{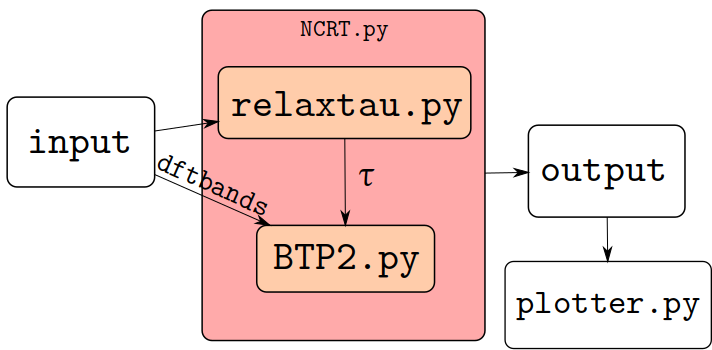}
\caption{Flux diagram of the code for NCRT.}
\label{fig9}
\end{figure}

The multiprocessing {\tt python} library is used for  $\tau$ and the Fermi integrals (currently only for {\tt ART.py}). 
The output is saved in {\tt json} format, and the data structure is based on  {\tt numpy} 
multidimensional arrays: for example, the figure of merit $ZT$ is a 3-dimensional array, the dimensions 
being cartesian component (with no magnetic field $ZT$ has only diagonal elements), the temperature 
$T$, and chemical potential $\mu$.

The operative structure comprises three  folders: {\tt script}, {\tt input} and {\tt data}. Folder {\tt script} contains the code in the three versions {\tt CRT.py}, {\tt ART.py}, {\tt NCRT.py}. Also, it hosts  the utilities  {\tt TAU.py}, which computes the relaxation times vs.\,$T$ and $E$ for plotting and external use, and {\tt plotter.py}, which plots a number of quantities of interest, as well as  the custom libraries {\tt funzVar.py} and {\tt relaxTau.py}. The latter contains the $\tau$ model, and can be edited  to include, for example, additional  scattering mechanisms. 
 
 Folder {\tt input} must contain two (optionally, three) files. The  file {\tt vasprun.xml} contains
 the ab initio  calculation results, including the band structure, obtained with  {\tt VASP} (for information on using other
DFT codes  refer to  the BT2 forum \cite{bt2forum}). The file {\tt input} contains control parameters (among which are the mesh multiplier of Sect.\ref{trcoe}, and the ranges of temperature, chemical potential, energy, etc.) and material parameters (dielectric permittivities, effective masses, deformation potentials, sound velocity,  LO phonon energies) that enter the relaxation time model, as well as the lattice thermal conductivity tensor. The latter can alternatively be read from the third optional file  {\tt latthcond}.

The chosen script (or scripts)  plus {\tt TAU.py}  are run in {\tt scripts}. The output, dumped in {\tt data}, is then analyzed using  {\tt plotter.py}, which reads and deserializes {\tt  json} data to {\tt numpy}  arrays, and   plots the  quantities of interest (and can of course be adapted to fit individual requirements). The package is stand-alone and its disk occupation is minor, so it  can be placed anywhere, for example  in a local directory for the material under study.  More information  is  in the package's {\tt README.pdf} file \cite{thermocasu} and in the Supplementary Material \cite{smat}. 

As already mentioned, NCRT is  fairly inexpensive   computationally, depending on the system and control parameters. For the present material, it costs a couple of hours  with about one Gbyte memory usage 
for production-level parameters. We will implement multiprocessing in NCRT too shortly, which will speed up the proceedings accordingly. ART, on the other  hand, will run in minutes, being lighter by its very nature as well as its  use of parallelism.

\section{Summary}

We  predicted a  monotonic  $ZT$ between 0.4 and 2.2 between  300 and  1200 K  under $n$-doping in the layered perovskite Sr$_2$Nb$_2$O$_7$ via  calculations of the electronic structure and transport coefficients. The optimal carrier density is in the low-10$^{20}$ cm$^{-3}$ range. At optimal density the Seebeck thermopower coefficient is between 220 and 320 $\mu$V/K, but can reach  800 $\mu$V/K at lower doping. The largest $ZT$ is along the $a$ crystal axis; other components are one to two orders of magnitude smaller. Much of the potential of this material is due to its small and almost $T$-independent lattice thermal conductivity.
We explored the use of constant or  energy- and temperature-dependent relaxation times. Averaged-time or full calculations are rather comparable in the present case, though of course this need not be the case in general. To help improve the treatment of relaxation time for other materials, we provide a code \cite{thermocasu} implementing our approach.
%
%


\begin{thebibliography}{99}
\bibitem{noi2}
R. Farris, M. B. Maccioni, A. Filippetti, and V. Fiorentini, J. Phys.: Condens. Matter {\bf 31}, 065702 (2019); M. B. Maccioni, R. Farris, and V. Fiorentini, Phys. Rev. B {\bf 98}, 220301(R) (2018).

\bibitem{LTO}
V. Fiorentini, R. Farris, E. Argiolas, and  M. B. Maccioni,
Phys. Rev. Materials {\bf 3}, 022401(R) (2019)


\bibitem{allen}
P. B. Allen, Phys. Rev. B {\bf  17}, 3725 (1978);
P. B. Allen, W. E. Pickett, and H. Krakauer, {\it ibid.} {\bf 37}, 7482 (1988).

\bibitem{thermocasu}
G. Casu, {\tt https://tinyurl.com/ycyvt4vk} 


\bibitem{vasp}
G. Kresse and J. Furthm\"uller, Phys. Rev. B {\bf 54}, 11169 (1996);
G. Kresse and D. Joubert, {\it ibid}. {\bf 59}, 1758 (1999).

\bibitem{bt2}
G. K. H. Madsen, J. Carrete, and M. J. Verstraete, Comp. Phys. Commun. {\bf 231}, 140 (2018).



\bibitem{SNO-therm}
 T. D. Sparks, P. A. Fuierer, and D. R. Clarkew, J. Am. Ceram. Soc. {\bf 93}, 1136 (2010). 
 
\bibitem{pbe}
J. P. Perdew, K. Burke, and M. Ernzerhof,
Phys. Rev. Lett. {\bf 77}, 3865 (1996).

 \bibitem{paw}
 P. E. Bl\"ochl, Phys. Rev. B {\bf {50}}, 17953 (1994); G. Kresse and D. Joubert, 
 {\it ibid.}, {\bf {59}}, 1758 (1999).
 
 
\bibitem{ohi}
S. Kojima, K. Ohi, M. Takashige, T. Nakamura, and H. Kakinuma
Solid State Commun. {\bf 31} 755 (1979)

 \bibitem{ohi2}
 Y. Akishige and  K. Ohi, Ferroelectrics {\bf 203}, 75 (1997).
 
  
  \bibitem{elda}
  Interestingly, we note in passing that   the calculated gap corrected with the empirical formula of 
  V. Fiorentini and A. Baldereschi, Phys. Rev. B {\bf 51}, 17196 (1995) with $\varepsilon_{\infty}$$\sim$6 \cite{epsinf} yields 4 eV as in experiment.
  
  
\bibitem{epsinf}  
A. Krause, W. M. Weber, D. Pohl, B. Rellinghaus, A. Kersch, and T. Mikolajick,
J. Phys. D: Appl. Phys. {\bf 48}, 415304 (2015) report $\varepsilon_{\infty}$$\simeq$6-6.5 for perovskites with similar electronic structure. Our  chosen value  overestimates  polar  scattering and produces slightly lower $ZT$. 

  \bibitem{ridley1}
B. K. Ridley, {\it Quantum Processes in Semiconductors} (Clarendon Press, Oxford, 1988).

\bibitem{ridley2}
B. K. Ridley, J. Phys.: Condens. Matter {\bf 10}, 6717 (1998).

\bibitem{smat}
See Supplemental Material at [URL inserted by publisher] for details on the scattering model and the code.

\bibitem{nye}
J. F. Nye, {\it Physical properties of crystals} (Clarendon Press, Oxford 1985).

\bibitem{electrom}
A. G. Kalinichev, J. D. Bass, C. S. Zhab, P. D. Han and D. A. Payne
 J. Appl. Phys. {\bf 74}, 6603 (1993);
 F. Chen, L. Kong, W. Song, C. Jiang, S. Tian, F. Yu, L. Qin, C. Wang, and X. Zhao,
 J. Materiomics, {\bf 5}, 73 (2019).
 
 
\bibitem{giustino}
F. Giustino,  Rev. Mod. Phys. {\bf 89}, 015003 (2017);
S. Ponc\'e, W. Li, S. Reichardt, and F. Giustino, Rep. Prog. Phys. {\bf 83}, 036501 (2020);
M. Fiorentini  and N. Bonini, Phys. Rev. B {\bf 94}, 085204 (2016).

\bibitem{cardona}
P. Y. Yu and M. Cardona, {\it Fundamentals of semiconductors} (Springer, Berlin 2010), Ch.5.
 
 \bibitem{epsil}
 {\tt https://doi.org/10.17188/1207555} .  
  
 \bibitem{discLO}
E Buixaderas, S Kamba, and J Petzelt, J. Phys.: Condens. Matter {\bf 13}, 2823 (2001).
The measured TO energy is rescaled  via the Lyddane-Sachs-Teller relation.


\bibitem{locq}
J. W. Seo, J. Fompeyrine, H. Siegwart, J.-P. Locquet, Phys. Rev. B 63, 205401 (2001).

\bibitem{ceram}
Yi-Cheng Liou, Wen-Chou Tsai, Jian-Yi Yu, and Hsiao-Chun Tsai, 
Ceramics International {\bf 41} 7036 (2015).
	
\bibitem{harm}
The harmonic average $x_h$ of {\bf x}=($x_1$, $x_2$, $x_3$) is
$$
\frac{3}{x_h}=\frac{1}{x_1} + \frac{1}{x_2} + \frac{1}{x_3}.
$$ 

\bibitem{bt2forum}
{\tt https://groups.google.com/forum/\#!forum/boltztrap}


\end{thebibliography}
\end{document}